\documentclass[aps,prl,nofootinbib,twocolumn,showpacs,preprintnumbers,superscriptaddress,letterpaper,amsmath,amssymb]{revtex4}
\usepackage{graphicx}  % needed for figures
\usepackage{dcolumn}   % needed for some tables
\usepackage{bm}        % for math
\usepackage{amssymb}   % for math
\usepackage{feynmf}%%%{feynmp}
\usepackage{slashed}
\usepackage{multirow}
\def\gsim{\lower0.5ex\hbox{$\:\buildrel >\over\sim\:$}}
\def\lsim{\lower0.5ex\hbox{$\:\buildrel <\over\sim\:$}}

\newcommand{\be}{\begin{equation}}
\newcommand{\ee}{\end{equation}}
\newcommand{\bea}{\begin{eqnarray}}
\newcommand{\eea}{\end{eqnarray}}

\newcommand{\nbox}{{\,\lower0.9pt\vbox{\hrule \hbox{\vrule height 0.2 cm
\hskip 0.2 cm \vrule height 0.2 cm}\hrule}\,}}

\def\missET {E_{\textrm{T}}^{\textrm{miss}}}
\def\tth {t\bar{t}H}

\begin{document}

\thispagestyle{empty}
\vspace*{-3.5cm}

\preprint{UCI-HEP-TR-2014-05}

\vspace{0.5in}

%\begin{flushright}
%\today\\
%\end{flushright}
%\vspace{0.5in}
\title{Erratum: Bounds on Invisible Higgs boson Decays from $t\bar{t}H$ Production}

\begin{center}
\begin{abstract}
Bounds on invisible decays of the Higgs boson from $t\bar{t}H$ production were inferred from a CMS search for stop quarks decaying to $t\bar{t}+\missET$.  Limits on the production of $t\bar{t}H$ relied on the efficiency of the CMS selection for $t\bar{t}H$, as measured in a simulated sample. An error in the generation of the simulated sample lead to a significant overestimate of the selection efficiency. Corrected results are presented.
\end{abstract}
\end{center}

\author{Ning Zhou}
\affiliation{Department of Physics and Astronomy, University of
  California, Irvine, CA 92697}
\author{Zepyoor Khechadoorian}
\affiliation{Department of Physics and Astronomy, University of
  California, Irvine, CA 92697}
\author{Daniel Whiteson}
\affiliation{Department of Physics and Astronomy, University of
  California, Irvine, CA 92697}
\author{Tim M.P. Tait}
\affiliation{Department of Physics and Astronomy, University of
  California, Irvine, CA 92697}

\maketitle

% introduction
%\linenumbers

In a recent Letter~\cite{Zhou:2014dba}, we recast a CMS search for supersummetric stop quarks ($\tilde{t}\rightarrow bW+\missET$)~\cite{Chatrchyan:2013xna} to set bounds on $t\bar{t}H$ production where the Higgs boson decays invisibly.  

\begin{figure}
\includegraphics[width=2in]{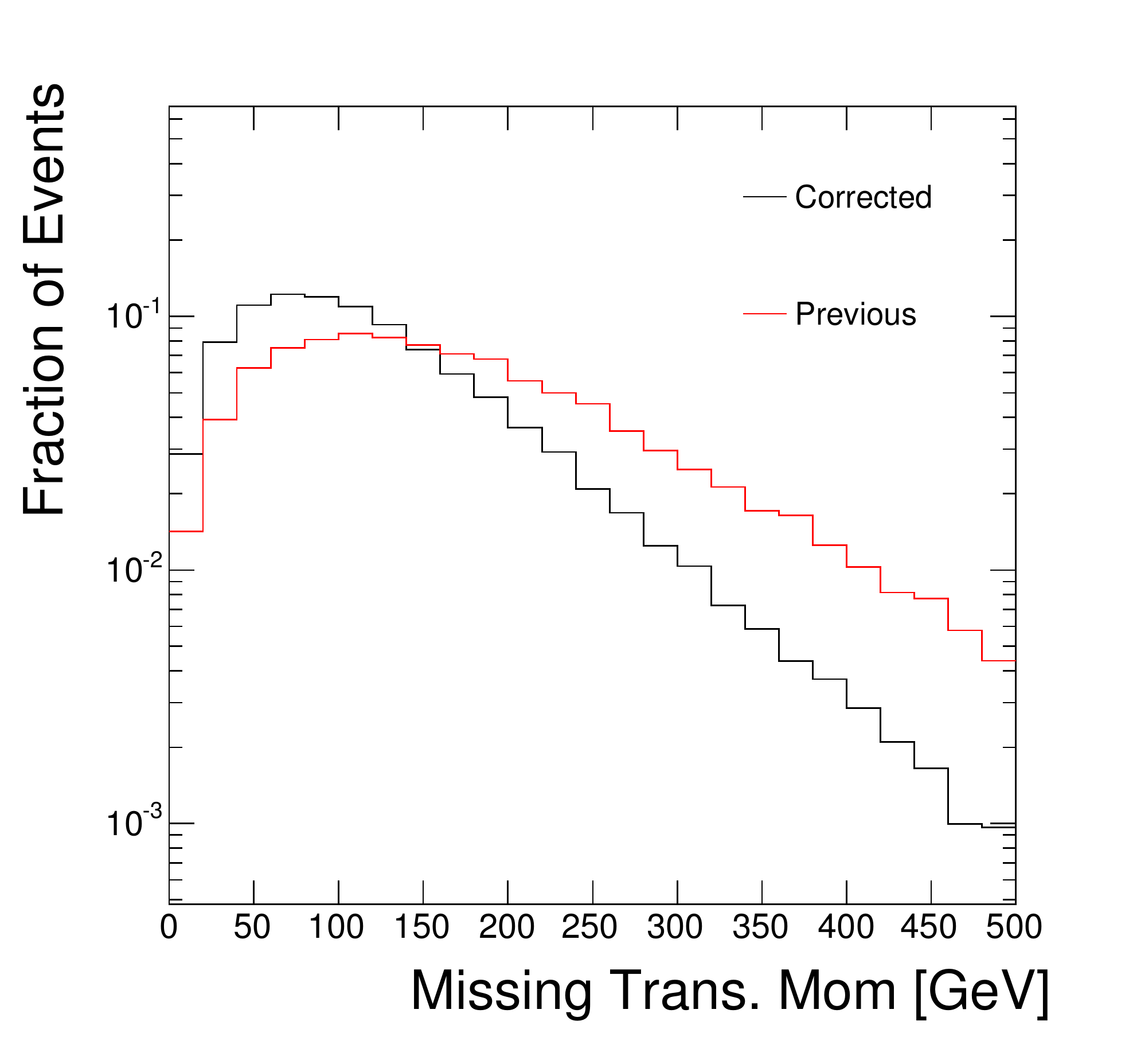}
\caption{ Distribution of missing transverse momentum $t\bar{t}H$ with $H\rightarrow$invisible for the simulated sample in the original paper as well as the corrected simulated sample.  Distributions are shown after requiring exactly one lepton, at least four jets and one $b$-tag. }
\label{fig:met}
\end{figure}

\begin{figure}
\includegraphics[width=1.5in]{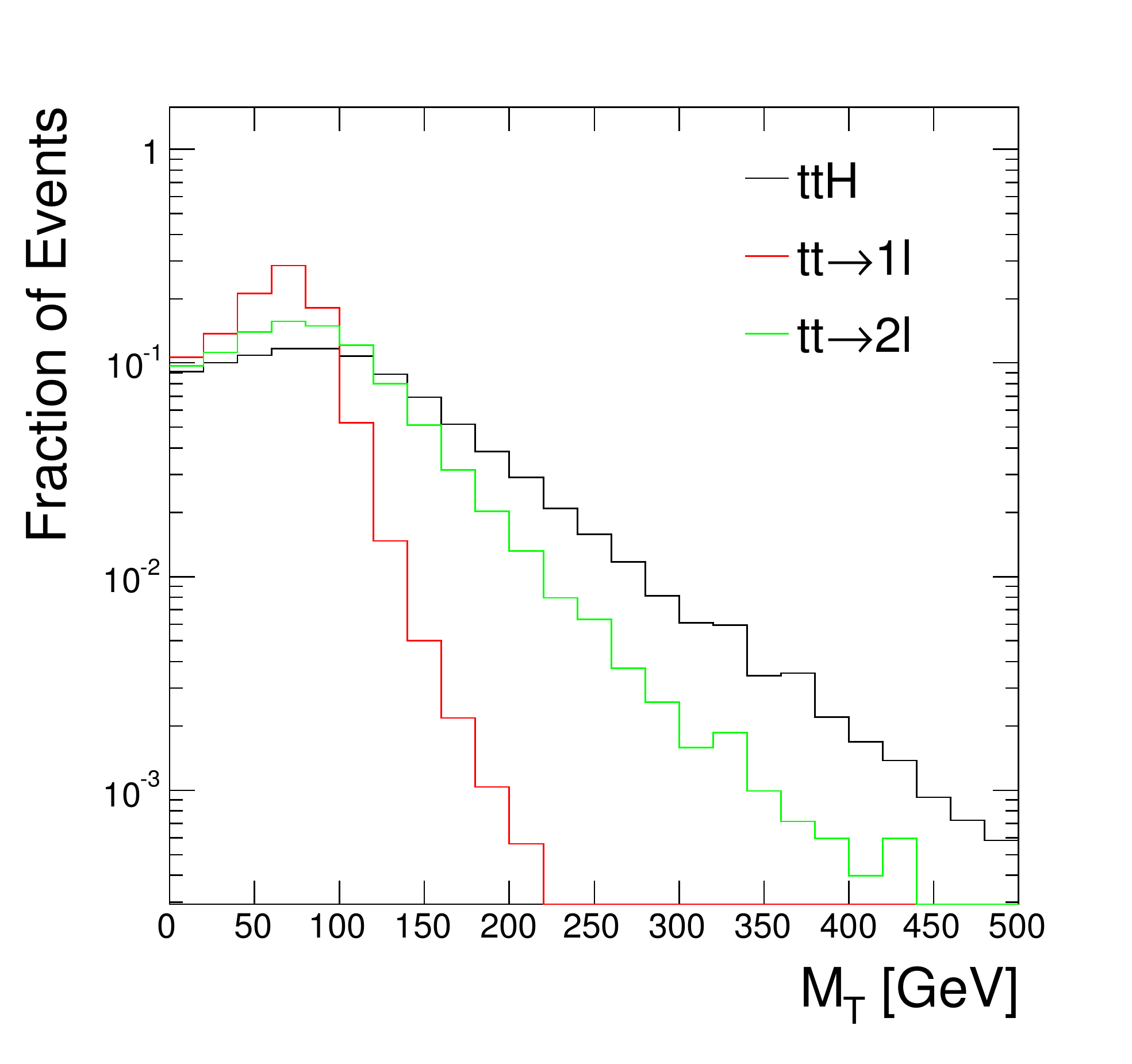}
\includegraphics[width=1.5in]{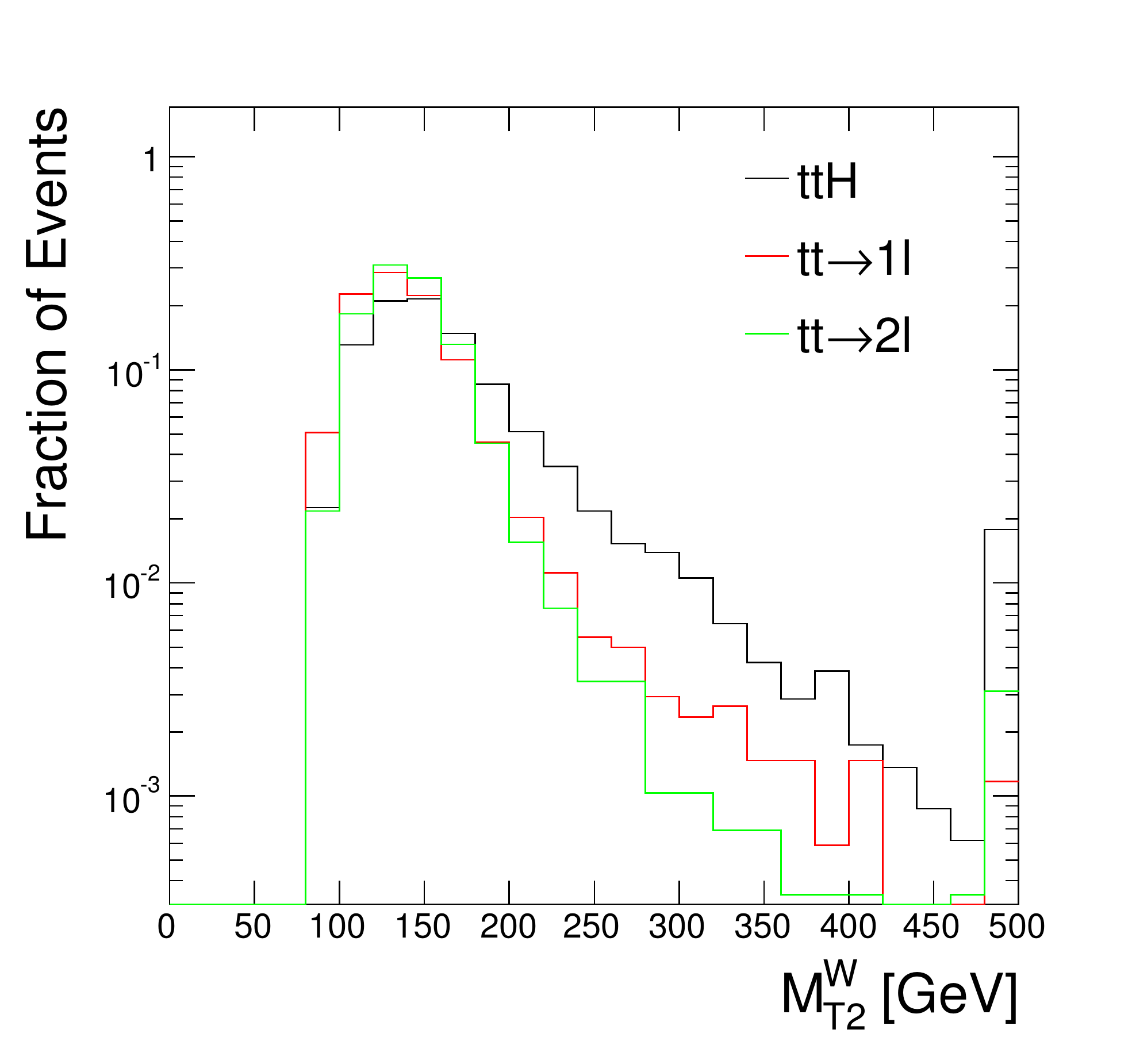}
\includegraphics[width=1.5in]{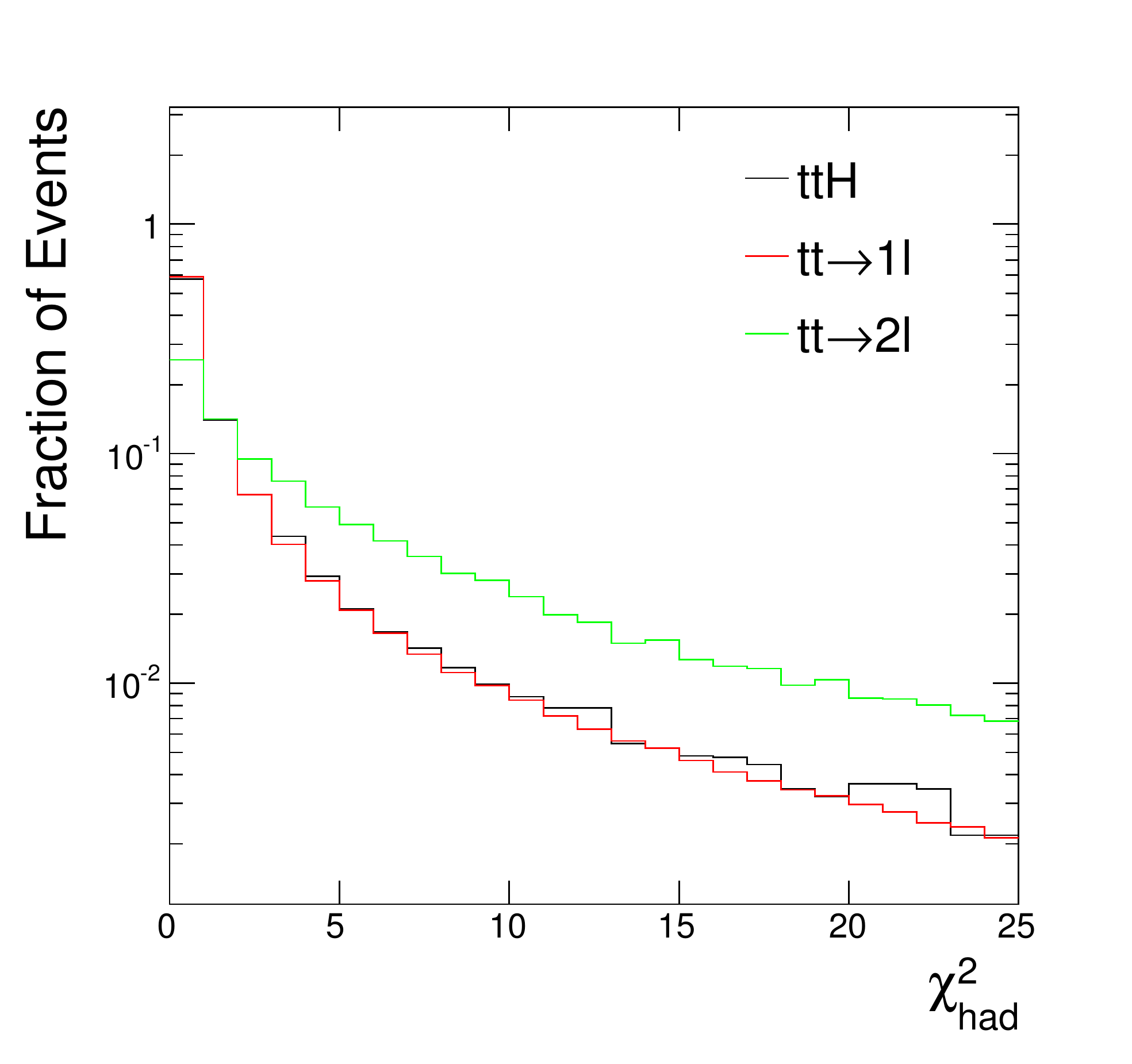}
\includegraphics[width=1.5in]{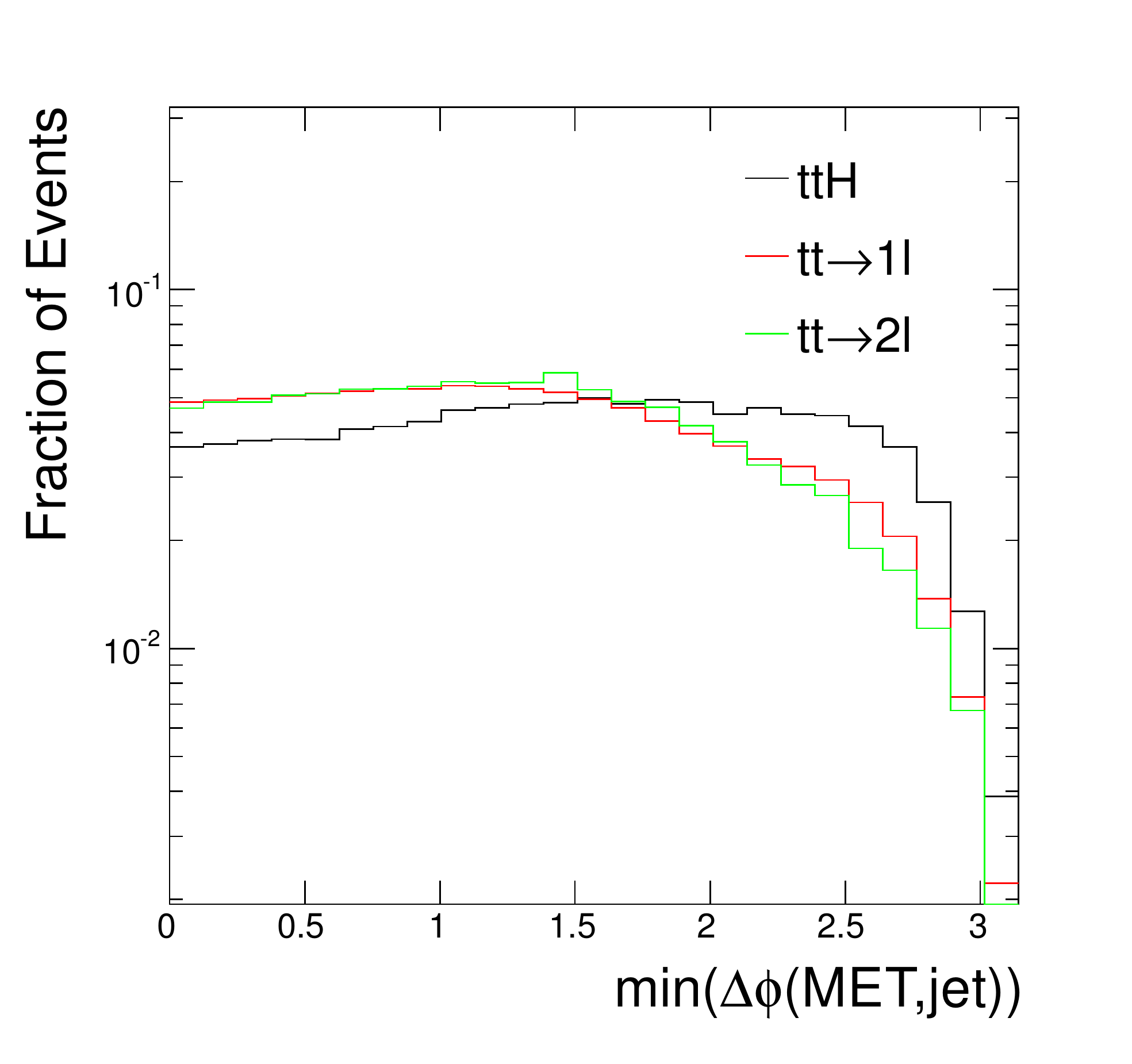}
\caption{Kinematics of $t\bar{t}H$ with $H\rightarrow$invisible and all $t\bar{t}$ decay modes, compared to the two dominant backgrounds, SM top quark production with either single lepton ($\ell\nu bqq'b$) or di-lepton ($\ell\nu b\ell'\nu b$) decay modes.  Distributions are shown of $M_T$, the transverse mass;  $M_{T2}^W$,  as defined in Ref.~\cite{Bai:2012gs}; $\chi_{\textrm{had}}^2$, the consistency of the $jjb$ system with a top quark hadronic decay; and min($\Delta\phi[$MET, jet]), the minimum angle between the $\missET$ and any jet . Distributions are shown after requiring exactly one lepton, at least four jets and one $b$-tag. }
\label{fig:kin}
\end{figure}

The result relies on the measurement of the efficiency of the CMS selection for $t\bar{t}H$ events, as measured in simulated samples generated with Madgraph5~\cite{Alwall:2011uj},  showering and hadronization with 
Pythia~\cite{Sjostrand:2006za} and detector response simulated by Delphes~\cite{deFavereau:2013fsa}.  The event generation used a model which included a $H\chi\bar{\chi}$ vertex to describe the invisible decay $H\rightarrow \chi\bar{\chi}$.  Inadvertantly, the contribution from diagrams containing a $ggH$ effective vertex was overestimated.  These diagrams tend to give larger transverse momentum to the Higgs boson, leading to larger measured $\missET$ and a significant overestimate of the selection efficiency.  Figure~\ref{fig:met} shows a comparison of the original flawed sample and a new, correct sample.  Figure~\ref{fig:kin} shows a comparison of the kinematics of the corrected $t\bar{t}H$ sample to the primary sources of background, top quark pair production.

For the corrected sample, we again apply the results of the CMS $\tilde{t}$ search to invisible Higgs boson decays by calculating the expected 
yield of $\tth$ in each of the signal regions. We calculate upper bounds on $\sigma(\tth)\times BF(H\rightarrow$inv.$)$ 
using a one-sided profile likelihood and the CLs technique~\cite{Read:2002hq,Junk:1999kv}, evaluated 
using the asymptotic approximation~\cite{Cowan:2010js}.  For each of the sixteen signal regions, we 
calculate the median expected limit on $\sigma(\tth)\times BF(H\rightarrow$inv.$)$. As with the original sample, the region with the 
strongest expected limit is that targeting $\tilde{t}\rightarrow t\tilde{\chi}$ in the high-$\Delta M$ regime, 
with $\missET>250$ GeV.  This region has the additional requirements of min($\Delta\phi[\missET,j])>0.8$,
$M_{T2}^W>200$~GeV and $\chi_{\textrm{had}}^2<5.0$. The expected background is reported to be $9.5 \pm 2.8$. With our simulated sample, we calculate an 
expected $\tth$ yield of 2.1 events if BF($H\rightarrow$ inv.$)=1.0$ (compare to 11.4 events for the incorrect sample).  The efficiency of this selection
for $\tth\rightarrow t\bar{t}\chi\bar{\chi}$ events with $m_H=125$ GeV is 0.085\% (compare to 0.45\% for the incorrect sample).

\begin{figure}
\includegraphics[width=2.5in]{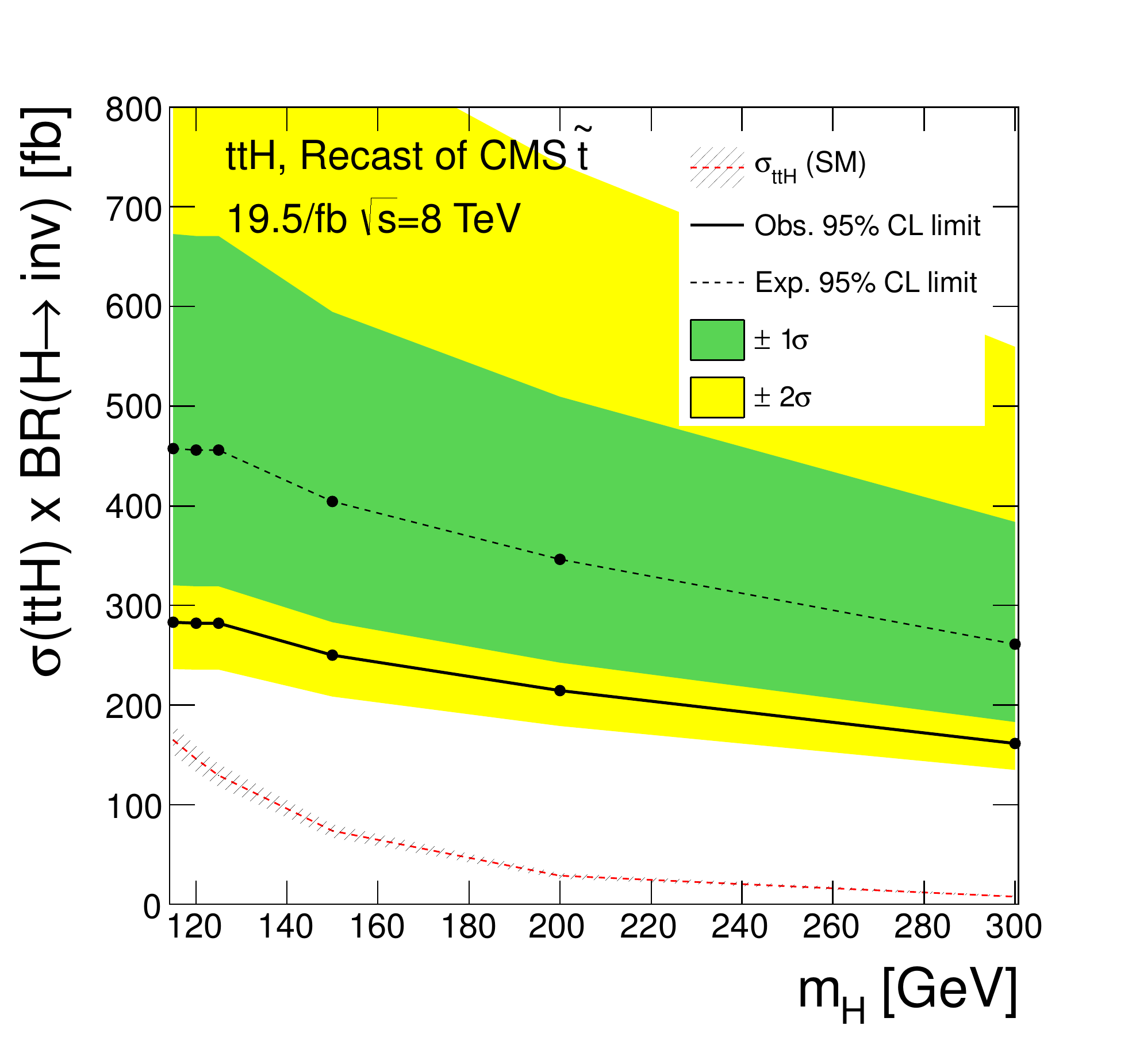}
\includegraphics[width=2.5in]{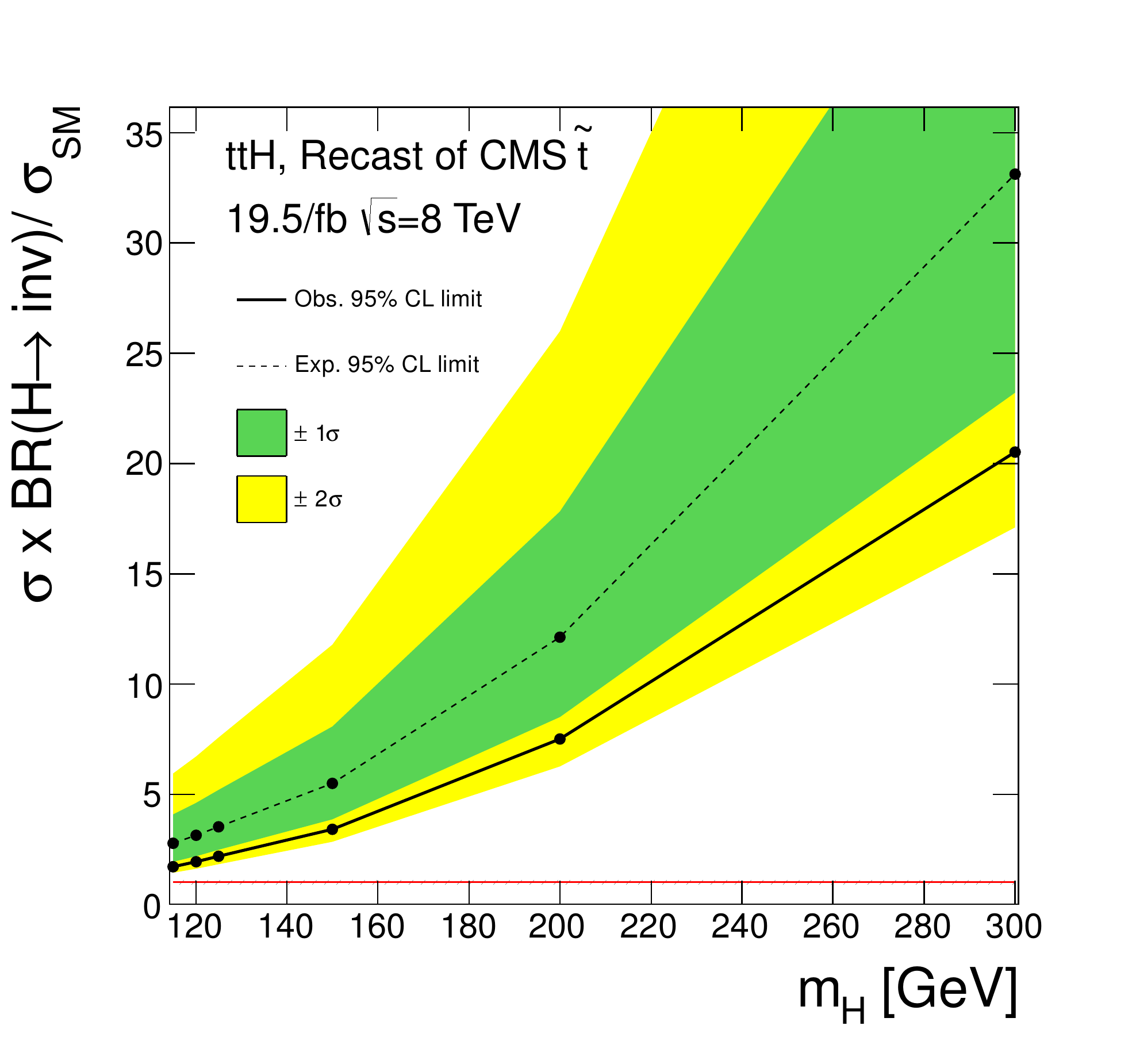}
\caption{ Top pane gives 95\% CL upper limits on $\sigma(t\bar{t}H) \times BF(H\rightarrow\textrm{inv.})$, including both expected and observed limits. Also shown is the SM rate of $\sigma(t\bar{t}H)$~\cite{Heinemeyer:2013tqa}.  The bottom pane shows the ratio of the constraint to the SM $\sigma(t\bar{t}H)$ cross section.}
\label{fig:lim}
\end{figure}

In this particular signal region, the data have fluctuated quite low, 
$N_{\textrm{obs}}=3$ events, giving an observed upper bound considerably stronger than the 
median expected results; see Fig.~\ref{fig:lim}.  Dividing by the predicted rate of $\tth$ production in the 
SM~\cite{Heinemeyer:2013tqa} gives a limit on BF($H\rightarrow$ inv$)$; the observed 
(expected) result is $<1.9~(3.0)$ at 95\% CL for $m_{H}=125$ GeV.

This result is significantly weaker than the previous. For this reason, we do not perform a combination with other channels nor provide an interpretation in terms of the Higgs portal model.

\bibliography{tth_err}

\end{document}